\documentclass[]{pasj01}

\begin{document} 
\Received{}
\Accepted{}

\title{Gas, Dust, Stars, Star Formation and their Evolution in M\ 33 at Giant Molecular Cloud Scales}

\author{Shinya \textsc{Komugi}\altaffilmark{1,2}}
\altaffiltext{1}{Department of Liberal Arts, Kogakuin University, 2665-1 Nakano-cho, Hachioji, Tokyo, Japan}
\altaffiltext{2}{Chile Observatory, National Astronomical Observatory of Japan, National Institutes of Natural Sciences, 2-21-1 Osawa, Mitaka, Tokyo 181-8588, Japan}
\email{skomugi@cc.kogakuin.ac.jp}

\author{Rie \textsc{E. Miura},\altaffilmark{2}}

\author{Nario \textsc{Kuno},\altaffilmark{3,4}}
\altaffiltext{3}{Division of Physics, Faculty of Pure and Applied Sciences, University of Tsukuba, 1-1-1 Tennodai, Tsukuba, Ibaraki 305-8571, Japan}
\altaffiltext{4}{Tomonaga Center for the History of the Universe, University of Tsukuba, 1-1-1 Tennodai, Tsukuba, Ibaraki 305-8571, Japan}

\author{Tomoka \textsc{Tosaki},\altaffilmark{5}}
\altaffiltext{5}{Joetsu University of Education, Yamayashiki-machi, Joetsu, Niigata 943-8512, Japan}


\maketitle

\begin{abstract}
We report on a multi parameter analysis of giant molecular clouds (GMCs) in the nearby spiral 
galaxy M33.  A catalog of GMCs identifed in $^{12}\mathrm{CO}(J=3-2)$ was used to compile 
associated $^{12}\mathrm{CO}(J=1-0)$, dust, stellar mass and star formation rate.  Each 
of the 58 GMCs are categorized by their evolutionary stage.  Applying the principal component analysis on these
parameters, we construct two principal components PC1 and PC2 which retain 75\% of the information in the original
dataset.  PC1 is interpreted as expressing the total interstellar matter content, and PC2 
as the total activity of star formation.  Young ($<$10Myr) GMCs occupy a distinct region in the PC1-PC2 
plane, with lower ISM content and star formation activity compared to intermediate age and older clouds.  
Comparison of average cloud properties in different evolutionary stages imply that GMCs may be heated or grow denser and more massive via aggregation of diffuse material in their first $\sim$10 Myr.  The PCA also
objectively identified a set of tight relations between ISM and star formation.  The ratio of the two CO lines
is nearly constant, but weakly modulated by massive star formation.  Dust is more 
strongly correlated with the star formation rate than the CO lines, supporting recent findings that 
dust may trace molecular gas better than CO.  Stellar mass contributes weakly to the 
star formation rate, reminiscent of an extended form of the Schmidt Kennicutt relation with the molecular gas term substituted by dust.
\end{abstract}

\section{Introduction}
The transition from gas to stars is a complex process involving various 
spatial and temporal scales.  Observationally, much of the star formation take place within 
giant molecular clouds (GMCs) which are typically few$\times 10$ to $\sim 100$ parsecs in 
size.  However, current knowledge of the general law of star formation is limited to semi-global
scales over kilo-parsecs, where the relation between molecular gas and star formation is 
characterized by a simple power law, the Schmidt-Kennicutt (SK) relation (\cite{kennicutt98,komugi05} and references therein).

The SK relation is known to break down at the scale of individual GMCs \citep{onodera10}
 when using diffuse molecular gas tracers like the $^{12}\mathrm{CO}(J=1-0)$ 
emission.  This has been attributed to the different evolutionary stage of GMCs 
(e.g., \cite{kawamura09}) which become evident at the $\sim$100 parsec scale \citep{onodera10}.  \citet{komugi12} support this by showing that the dispersion of the SK relation becomes small when 
star forming regions at the same age are used.  When using denser and/or warmer gas such as that traced by 
the $^{12}\mathrm{CO}(J=3-2)$ line, the relation is generally tighter \citep{komugi07,muraoka07,iono09,muraoka16}
and this has been explained by the higher CO transition tracing molecular gas that is spatially 
and temporally closer to the actual site of star formation (c.f., \cite{wu05} using HCN emission).  
\citet{onodera12} reveal a connection between the two CO tracers and star formation, finding that 
larger GMCs tend to have higher $^{12}\mathrm{CO}(J=3-2)$/$^{12}\mathrm{CO}(J=1-0)$ ratios ($r_{31}$), 
along with a higher star formation rate ($SFR$).

The connection of dust and star formation is also important, as dust grain surfaces
act as a catalysis for molecular gas production.  Recent studies \citep{wolfire10, paradis12, scoville16} also indicate that dust may trace molecular gas better 
than CO in some regions where gas is not dense enough to shield itself against UV photon 
dissociation \citep{hollenbach99}.
Stellar mass may be another parameter which controls star formation at various scales.  \citet{shi11} and \citet{rahmani16} show that when the SK relation is extended to include the stellar mass, the dispersion becomes smaller both globally and within galaxies at kilo-parsec scales.  Stellar contribution is also theoretically expected from 
stability analysis \citep{dib17} and pressure regulated star formation (\cite{blitz06} and others).

It is becoming increasingly clear that at the scale of GMCs, star formation is not explained by a single parameter
equation such as the SK relation; it is better described as a time-evolving equilibrium state 
between ISM and star formation.  In this paper, we
attempt to formalize this state using a multi-parameter analysis on a sample of GMCs
in the nearby spiral galaxy M33.

\section{Data}

The Nobeyama Radio Observatory (NRO) M33 All-Disk Survey of Giant Molecular Clouds (MAGiC) 
\citep{kuno11} observed the optical disk ($30^{\prime}\times 30^{\prime}$, corresponding to 7.3 kpc $\times$ 
7.3 kpc) in the nearby spiral galaxy M33 using the $^{12}\mathrm{CO}(J=1-0)$ line at 
$19^{\prime \prime}$ resolution, corresponding to 80 pc \citep{tosaki11}.  A corresponding map was obtained at ASTE 
telescope \citep{ezawa04,kohno05} in $^{12}\mathrm{CO}(J=3-2)$ at $25^{\prime \prime}$ (100 pc) resolution
by \citet{miura12}, and in the 1.1\ mm continuum by \citet{komugi11m33} at $40^{\prime \prime}$ (160 pc) 
resolution. In this paper, we use $^{12}\mathrm{CO}(J=3-2)$, $^{12}\mathrm{CO}(J=1-0)$ and $SFR$
derived from combinations of H$\alpha$ and $24 \mu \mathrm{m}$, for 71 GMCs catalogued by \citet{miura12}.
In order to circumvent the uncertainties associated with the CO to $\mathrm{H_2}$ conversion factor, 
throughout this paper we use CO luminosities $L'_{\mathrm{3-2}}$ and $L'_{\mathrm{1-0}}$ instead of converting them to molecular gas masses.

The catalog by \citet{miura12} also includes a measure of the evolutionary stage of the GMCs, 
Type A through D, based on stellar group identifications using the \citet{massey06}
star catalog and Padova stellar synthesis tracks \citep{marigo08,girardi10}.  In this paper, we use only Type
B, C, and D GMCs, because only 1 Type A GMC was identified in the catalog, and further it was not detected at 
1.1\ mm.  Type B are young GMCs which are associated with relatively small HII regions but not with young 
stellar groups (YSGs), and are at a stage approximately 3-7 Myr after the first formation of massive OB stars.  
Type C GMCs are associated with both HII regions and YSGs less than 10 Myr old, indicating that they are 10-20 
Myr in age.  Type D GMCs have HII regions and relatively old (10-30 Myr) stellar groups, and are 20-40 Myr in 
age (see \cite{miura12}).

We identified cold dust clumps associated with each of these catalogued GMCs using the 
1.1\ mm continuum map from \citet{komugi11m33}.  A box of $54^{\prime \prime}$ on the side, 
which is two times the detector FWHM of the observation, was centered on the HII region to search for
a 1.1\ mm peak.  The angular size of the 1.1\ mm beam 
covers the entire GMC, so its peak value can be used to obtain the total dust mass.
We assume a dust emissivity of $\kappa = 0.114\ \mathrm{m^2 kg^{-1}}$, and corrected for the
dust temperature using the color temperature between Spitzer MIPS 160 $\mu \mathrm{m}$.  
Details of the 1.1 mm data and temperature derivation are found in \citet{komugi11m33}.

The stellar mass $M_*$ of the corresponding regions were derived from $K_\mathrm{S}$ (2.1$\mu \mathrm{m}$)
image from the 2MASS Large Galaxy Survey \citep{jarrett03}.  The image was background subtracted using planar
interpolation from regions well away from the galaxy, and then convolved to $40^{\prime \prime}$ resolution
to match the 1.1mm image.  A Galactic extinction correction of 0.022 mag. was applied, based on the NASA 
Extragalactic Database (NED).  The $K_\mathrm{S}$ brightness was then converted to stellar mass using relations in \citet{blitz06}
assuming galactic inclination of $55^{\degree}$ \citep{vaucouleurs59}.

Peaks identified in 1.1mm and $K_\mathrm{S}$ were both inspected by eye to ensure that they are discrete 
clumps and are associated to the HII region and GMCs identified in both CO lines.  Out of the 71 regions catalogued 
in \citet{miura12}, 58 GMCs were identified in all of the two CO lines, $M_\mathrm{d}$, $SFR$ and $M_*$.  The final sample consists of 10 Type B, 30 Type C, and 18 Type D GMCs.  All following analyses in this paper use this final sample.

\section{Principal Component Analysis}

We utilize the Principal Component Analysis (PCA) approach to the catalogue of GMCs.
The PCA works on an arbitrary $n$ number of parameters $x$, and constructs $n$ new axes (principal components; PCs) using a linear combination of these $n$ parameters, so that the new axes ``better describe'' the characteristics of the sample than the original parameters.  Formally, 
\begin{equation}
\mathrm{PC}_i = \alpha_{i,1} x_1 + \alpha_{i,2} x_2 + ... + \alpha_{i,n} x_n
\end{equation}
where $x_i$ are explanatory parameters of the GMC samples, which are $\log SFR$, $\log L'_\mathrm{3-2}$, $\log L'_\mathrm{1-0}$, $\log M_{*}$ and $\log M_\mathrm{d}$.  
Since the parameters $x_i$ have different dynamical range and units, they are normalized so that their average is zero and variance is unity.
The coefficients $\alpha$ are normalized so that
\begin{equation}
\sqrt{\alpha^2_{i,1} + \alpha^2_{i,2} + ... + \alpha^2_{i,n}} = 1.
\end{equation}
The amount of information that is projected onto the new PC can be measured by their variance (eigen value).  In the present study, $n=5$, and the total variance of 5 is conserved after the projection of the axes.  The coefficients $\alpha$ represent the amount of contribution that each parameter $x_i$ has on $\mathrm{PC}_i$.
 The first principal component, $\mathrm{PC_1}$, is defined so that the variance of data around this axis is maximized.
$\mathrm{PC_2}$ is then defined orthogonal to $\mathrm{PC_1}$, having the next largest variance, and so on.  The $n$-th principle component PC$_n$ will have the minimum variance.  In short, $\mathrm{PC_1}$ is a parameter which best describes the GMC sample, and $\mathrm{PC_2}$ follows.

Although molecular clouds are characterized by their CO luminosities and their associated dust, stellar mass and $SFR$, the SK relation is usually described in terms of surface densities of each quantity.  We have applied the PCA in two ways, to both luminosity (mass) and surface densities.  The surface density of each parameters were calculated by dividing each quantity by the projected area of the molecular cloud as derived in \citet{miura12}, and the stellar surface density was derived directly from the $K_\mathrm{S}$ brightness.

\section{Results}

Tables 1 and 2 summarize the variances determined for the PCs.  For both cases, $\mathrm{PC_1}$, $\mathrm{PC_2}$ and $\mathrm{PC_3}$ have variance adding up to more than $90\%$ of the total information in the original dataset.
For $\mathrm{PC_4}$ and $\mathrm{PC_5}$, the amount of information is significantly reduced from the original axes; the variance 
around these PCs decrease from 1 to 0.42 (0.29) and 0.07 (0.07), corresponding to only 8\% (6\%) and 1\% (1\%) of the total variance, respectively, where values in parentheses are for surface densities.  
Tables \ref{coeff_lum} and \ref{coeff_surface} summarize the coefficients $\alpha$.  
The PCA is a purely mathematical operation, so a physical interpretation must be given to the new PCs.
  For a robust interpretation, we adopt only the terms that contribute significantly to the new PCs, and ignore terms with coefficients smaller than 0.45 (corresponding to $\alpha^2=1/n$ in equation 2 for $n=5$).  Coefficients which contribute significantly to the PCs are simliar for luminosities and surface densities, except for the case of PC3.  They are shown by boldface in tables \ref{coeff_lum} and \ref{coeff_surface}.

\begin{table*}[ht]
\caption{Variance of Principle Components for Luminosity}\label{pca_lum}
\begin{center}
\begin{tabular}{lccccc}
\hline
values & $\mathrm{PC_1}$ & $\mathrm{PC_2}$ & $\mathrm{PC_3}$ & $\mathrm{PC_4}$ & $\mathrm{PC_5}$\\
\hline
 Eigen Value (variance)  & $2.68\pm 0.14$  & $1.23\pm 0.02$ & $0.81\pm 0.01$ & $0.42\pm 0.00$  & $0.07\pm 0.00$  \\
 Proportion of variance & $0.52\pm 0.01$  & $0.24\pm 0.01$  & $0.16\pm 0.00$ & $0.08\pm 0.00$  & $0.01\pm 0.00$  \\
 Cumulative Proportion  & 0.52 & 0.75 & 0.91 & 0.99 & 1.00  \\
\hline
\end{tabular}\\
\end{center}
\begin{tabnote}
The eigen values of each of the PCs correspond to the amount of information that is projected onto the PC.  The propertion of variance is its ratio to the total amount of information, i.e., the variance divided by $n=5$.
\end{tabnote}
\end{table*}
\begin{table*}[ht]
\caption{Variance of Principle Components for Surface Density}\label{pca_surface}
\begin{center}
\begin{tabular}{lccccc}
\hline
values & $\mathrm{PC_1}$ & $\mathrm{PC_2}$ & $\mathrm{PC_3}$ & $\mathrm{PC_4}$ & $\mathrm{PC_5}$\\
\hline
 Eigen Value (variance)  & $2.64\pm 0.13$ & $1.30\pm 0.03$ & $0.91\pm 0.02$ & $0.29\pm 0.00$ & $0.07\pm 0.00$ \\
 Proportion of variance &  $0.51\pm 0.01$ & $0.25\pm 0.01$ & $0.17\pm 0.01$ & $0.06\pm 0.00$ & $0.01\pm 0.00$ \\
 Cumulative Proportion  & 0.50 & 0.76 & 0.93 & 0.99 & 1.00 \\
\hline
\end{tabular}\\
\end{center}
\begin{tabnote}
Same as table 1, but for surface densities.
\end{tabnote}
\end{table*}

\begin{table*}[ht]
\caption{PCA Coefficients $\alpha$ for Luminosities}\label{coeff_lum}
\begin{center}
\begin{tabular}{lccccc}
\hline
original parameters & $\mathrm{PC_1}$ & $\mathrm{PC_2}$ & $\mathrm{PC_3}$ & $\mathrm{PC_4}$ & $\mathrm{PC_5}$\\
\hline
$\log L'_\mathrm{1-0}$ & $\mathbf{0.50 \pm 0.02}$  & $\mathbf{-0.49 \pm 0.02}$   & $-0.26\pm 0.01$               & $0.06\pm 0.00$                & $\mathbf{-0.67 \pm 0.00}$ \\
$\log L'_\mathrm{3-2}$ & $\mathbf{0.56 \pm 0.01}$  & $-0.34\pm 0.02$                  & $-0.10\pm 0.01$               & $0.21\pm 0.01$                 & $\mathbf{0.72 \pm 0.00}$ \\
$\log M_\mathrm{d}$    & $\mathbf{0.49 \pm 0.01}$  & $0.23\pm 0.03$                   & $0.32\pm 0.02$                & $\mathbf{-0.78\pm 0.00}$  & $0.00\pm 0.00$\\
$\log M*$                      & $0.21\pm 0.03$                  & $\mathbf{0.60\pm 0.02}$    & $\mathbf{-0.77\pm 0.02}$   & $0.00\pm 0.02$               & $0.01\pm 0.00$\\
$\log SFR$                   & $0.40\pm 0.02$                  & $\mathbf{0.48\pm 0.02}$    & $\mathbf{0.48\pm 0.02}$    & $\mathbf{0.59\pm 0.00}$ & $-0.19\pm 0.00$  \\
\hline
\end{tabular}
\end{center}
\end{table*}
\begin{table*}[ht]
\caption{PCA Coefficients $\alpha$ for Surface Densities}\label{coeff_surface}
\begin{center}
\begin{tabular}{lccccc}
\hline
 original parameters & $\mathrm{PC_1}$ & $\mathrm{PC_2}$ & $\mathrm{PC_3}$ & $\mathrm{PC_4}$ & $\mathrm{PC_5}$\\
  \hline
$\log L'_\mathrm{1-0}$ & $\mathbf{0.49\pm 0.02}$ & $\mathbf{-0.45\pm 0.01}$  & $-0.35\pm 0.02$            & $-0.02\pm 0.00$                & $\mathbf{-0.66\pm 0.00}$ \\
$\log L'_\mathrm{3-2}$ & $\mathbf{0.57\pm 0.01}$ & $-0.28\pm 0.02$              & $-0.24\pm 0.01$              & $0.08\pm 0.01$                 & $\mathbf{0.73\pm 0.00}$ \\
$\log M_\mathrm{d}$    & $\mathbf{0.48\pm 0.02}$ & $0.24\pm 0.03$               & $\mathbf{0.51\pm 0.02}$  & $\mathbf{-0.68\pm 0.00}$ & $0.05\pm 0.00$\\
$\log M*$                      & $0.07\pm 0.03$             & $\mathbf{0.65\pm 0.03}$   & $\mathbf{-0.71\pm 0.03}$ & $-0.26\pm 0.01$               & $0.01\pm 0.01$\\
$\log SFR$                   & $0.44\pm 0.02$              & $\mathbf{0.49\pm 0.03}$   & $0.25\pm 0.03$               & $\mathbf{0.69\pm 0.01}$   & $-0.16\pm 0.00$ \\
\hline
\end{tabular}
\end{center}
\end{table*}

\subsection{Robustness of PCA Analysis}
The PCA analysis is sensitive to uncertainties in the original dataset.  Therefore, we estimate the uncertainty in the variances, PCA scores and coefficients by using a bootstrapping method .  We create a resampled dataset where values for each original parameter for each GMC follow a gaussian distribution with 30\% standard deviation with respect to the original data.  The assumed uncertainty of 30\% corresponds to typical error in the $L'_\mathrm{3-2}$ measurement, which has the largest observational error in the datasets, and therefore gives a conservative estimate of how the PCA analysis can be affected by intrinsic uncertainties in the dataset.  We create $10^4$ such resampled datasets, and run the PCA analysis on each of them.  The error estimates in figure 1 and tables 1 through 4 correspond to the standard deviation of each values computed from bootstrapping.  The errors for the variances, coefficients and PCA scores show that the PCA analyses are stable and results presented here are not severely affected by observational uncertainties.

\subsection{PC1 and PC2}

The first two PCs contain the most information in the dataset.  $\mathrm{PC_1}$ has the largest variance, 
contributing $\sim 50\%$ to the total value, thus best characterizes the star forming GMCs.  From tables \ref{coeff_lum} and \ref{coeff_surface}, $\mathrm{PC_1}$ is contributed almost equally by $L'_\mathrm{1-0}$, 
$L'_\mathrm{3-2}$ and $M_\mathrm{d}$ which all express quantities of the ISM, all with positive coefficients.  $\mathrm{PC_1}$ may be interpreted as a
scaling factor indicating the \textit{total ISM content} of the regions.
$\mathrm{PC_2}$ correlates positively with $SFR$ and stellar mass, both of which are related to star formation in the ISM.  A higher $SFR$, and its integration over time (the stellar mass) lead 
to a higher $\mathrm{PC_2}$.  It is also anti-correlated with $L'_\mathrm{1-0}$, which indicates that $\mathrm{PC_2}$ increases for a higher $SFR$ \textit{per molecular gas mass}, i.e., the efficiency of star formation.  $\mathrm{PC_2}$ may be interpreted as a measure of the \textit{star formation activity}, both instantaneous and over a longer timescale.  The first two PCs combined account for $\sim$75\% of total information.  They can be written as 
\begin{eqnarray}
\mathrm{PC^{\prime}_1} & = & 0.50\pm 0.02\ (0.49\pm 0.02) \tilde{\log L'_\mathrm{1-0}} \nonumber \\
& + & 0.56\pm 0.01\ (0.57\pm 0.01) \tilde{\log L'_\mathrm{3-2}} \nonumber \\
& + & 0.49\pm 0.01\ (0.48\pm 0.02)\tilde{\log M_\mathrm{d}}
\end{eqnarray}
\begin{eqnarray}
\mathrm{PC^{\prime}_2} & = & 0.60\pm 0.02\ (0.65\pm 0.03) \tilde{\log M_*} \nonumber \\
& + & 0.48\pm 0.02\ (0.49\pm 0.03) \tilde{\log SFR} \nonumber \\
& - & 0.49\pm 0.02\ (-0.45\pm 0.01) \tilde{\log L'_\mathrm{1-0}}
\end{eqnarray}
where the prime on the left hand side indicate that insignificant terms on the right hand side have been omitted, and tilde on the right hand side indicate that the (logarithm of) values are normalized as explained in section 3.  Values in parentheses indicate the case for surface densities (same hereafter).  Figure 1 shows $\mathrm{PC_1}$ versus $\mathrm{PC_2}$, categorized by their evolutionary stages.  For both luminosity and surface density, young Type B molecular clouds are distributed in a different region from intermediate (Type C) and older (Type D) clouds, such that they have less ISM content and star formation activity.

\begin{figure*}[tb]\label{fig_pca}
\begin{center}
\begin{tabular}{c}
  \begin{minipage}{0.5\hsize}
    \begin{center}
    \includegraphics[clip,width=8cm]{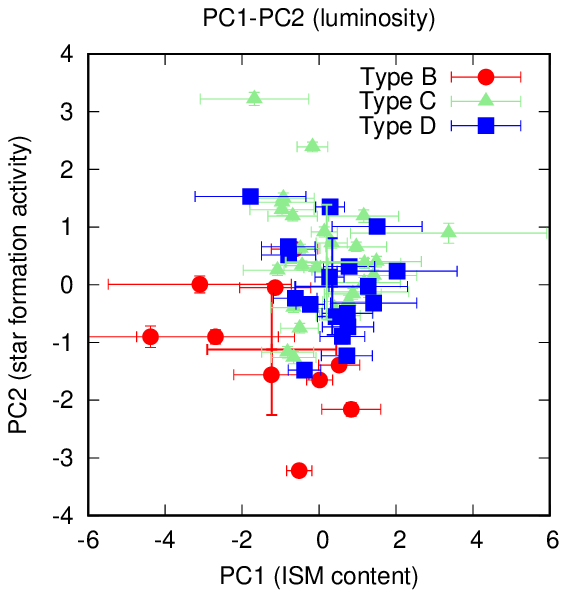}\label{pca_lum}
    \end{center}
  \end{minipage}
  \begin{minipage}{0.5\hsize}
    \begin{center}
    \includegraphics[clip,width=8cm]{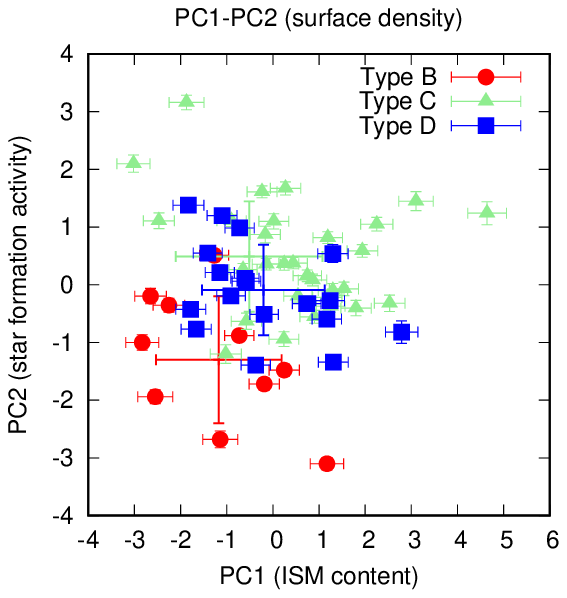}\label{pca_den}
    \end{center}
  \end{minipage}
 \end{tabular}
\caption{The total ISM content (PC1) and star formation activity (PC2) in M33, when considering luminosities (left) and surface densities (right).  For both panels, red circles indicate young Type B GMCs, green triangles indicate intermediate Type C GMCs, and blue squares indicate old Type D GMCs (see section 2 for their corresponding ages).  The crosses indicate the average and standard deviation for each GMC group.}
\end{center}
\end{figure*}

\subsection{PC3}
The third PC has a variance that is lower than unity.  This indicates that the amount of information has decreased from the original axes, i.e., that $\mathrm{PC_3}$ does not explain the properties of molecular clouds better than the original parameters.  It is dominated by contribution from the stellar mass, with smaller contribution from SFR in the case of luminosities, and dust in the case of surface densities.  The physical interpretation of $\mathrm{PC_3}$ is unclear, as the contributing parameters are different between luminosities and surface densities.  Here we will not attempt to give $\mathrm{PC_3}$ a physical interpretation.

\subsection{PC4 and PC5}
The last two PCs have the least variance in the given datasets, and can be used to detect relations between variables.  
Since the PC scores of individual clouds average to zero, we can write
\begin{eqnarray}
\mathrm{PC}^{\prime}_4 & = & 0.59\pm 0.00\ (0.69\pm 0.01) \tilde{\log SFR}  \nonumber \\
& - & 0.78\pm 0.00\ (-0.68\pm 0.00)\tilde{\log M_\mathrm{d}} = 0
\end{eqnarray}
\begin{eqnarray}
\mathrm{PC}^{\prime}_5 & = & 0.72\pm 0.00\ (0.73\pm 0.00)\tilde{\log L'_\mathrm{3-2}} \nonumber \\
& - & 0.67\pm 0.00\ (-0.66\pm 0.00) \tilde{\log L'_\mathrm{1-0}} = 0
\end{eqnarray}
Equations 5 and 6 indicate that there are two independent correlations in this dataset, one between $\log M_\mathrm{d}$ and $\log SFR$, and another between $\log L'_\mathrm{3-2}$ and $\log L'_\mathrm{1-0}$.  The relation between the original parameters can be derived by converting equations 5 and 6
to untilded values using the average and standard deviation of each parameters.  These averages and deviations are given in table \ref{avg_param}.  Althernatively, a simple least squares fitting to the original parameters (in which case the PCA is used only for selection of important parameters) can be used for comparison with the literature.  The two procedures give

\begin{eqnarray}
\log SFR & = & 3.4\log M_\mathrm{d} - 12.5   \quad (\mathrm{PC}^{\prime}_4 =0) \\
\log SFR & = & (1.4 \pm 0.3)\log M_\mathrm{d} - (7.1 \pm 0.8) \nonumber \quad \mathrm{(least\ squares\ fitting)}
\end{eqnarray}
and
\begin{eqnarray}
\log L'_\mathrm{3-2} & = & 0.95\log L'_\mathrm{1-0} + 0.65  \quad (\mathrm{PC}^{\prime}_5 =0) \\
\log L'_\mathrm{3-2} & = & (0.97 \pm 0.06)\log L'_\mathrm{1-0} + 0.7\pm 0.3 \nonumber \quad \mathrm{(least\ squares\ fitting)}
\end{eqnarray}
for luminosity, and 
\begin{eqnarray}
\log \Sigma_{SFR} & = & 3.1\log \Sigma_{\mathrm{d}} - 2.1 \quad (\mathrm{PC}^{\prime}_4 =0)  \\
\log \Sigma_{SFR} & = & (2.1 \pm 0.3)\log \Sigma_{\mathrm{d}} - (5.1 \pm 0.3) \nonumber \quad \mathrm{(least\ squares\ fitting)}
\end{eqnarray}
and
\begin{eqnarray}
\log \Sigma_{L'(3-2)} & = & 1.00\log \Sigma_{L'(1-0)} + 0.59  \quad (\mathrm{PC}^{\prime}_5 =0) \\
\log \Sigma_{L'(3-2)} & = & (0.99 \pm 0.06)\log \Sigma_{L'(1-0)} +(0.60 \pm 0.04) \nonumber  \quad \mathrm{(least\ squares\ fitting)}
\end{eqnarray}
for surface density.  Equation 7 through 10 indicate that a classical SK relation is not found between individual molecular clouds, but instead exposes the correlation between $SFR$ and dust.  Figure 2 shows equation 7 and 9, along with a SK relation plot comparing the $SFR$ with dust and the two CO lines.

\section{Discussion}
\subsection{Molecular gas and star formation at GMC scales}
A correlation between molecular gas and $SFR$ is non-existent when molecular gas is traced by the $^{12}\mathrm{CO}(J=1-0)$ line (figure 2 top right and bottom right with correlation coefficient 0.18 and 0.20, respectively) but slightly more correlated when using higher density tracers like the $^{12}\mathrm{CO}(J=3-2)$ (figure 2 top middle and bottom middle with correlation coefficient 0.40 and 0.42, respectively).  This is consistent with previous studies which find that molecular lines which trace denser or warmer molecular gas is more strongly correlated with star formation \citep{komugi07,muraoka07,iono09,muraoka16}, presumably because they trace gas that is spatially and temporally closer to star forming molecular gas.
The correlation between $SFR$ and dust mass is much more pronounced (equations 7 and 9, figure 2 top left and bottom left with correlation coefficient 0.64 and 0.66, respectively) than the relation for CO.  If we assume that molecular gas is the direct ingredient of star formation, and that their quantities are intrinsically correlated, the strong dust-SFR correlation supports recent studies which claim that dust emission traces molecular gas better than CO \citep{wolfire10,paradis12, scoville16}.  The basis behind this may be the existence of ``CO-dark'' molecular gas, in the diffuse outskirts of the molecular clouds where CO can be dissociated by UV photons but hydrogen molecules survive due to self-shielding \citep{hollenbach99,wolfire10}.  In this case, the strong correlation between $SFR$ and dust may be specific to M\ 33 because of its relatively low metallicity ($12+\log (O/H) = 8.48$ in the central region; \cite{bresolin11}), since the amount of CO-dark gas is inferred to increase in low metallicity systems \citep{israel97, leroy07, komugi11b}.  Alternatively, a strong correlation between dust and $SFR$ could imply that the 1.1\ mm observation is predominantly tracing dense gas that is linked directly to star formation, since millimeter dust is optically thin where CO would be optically thick and unable to accurately trace gas quantity.

\subsection{GMC evolution}
The distribution of GMCs in figure 1 has implications for how GMCs evolve after stars form.  Directly after the onset of star formation represented by Type B clouds, both the total ISM content ($\mathrm{PC_1}$) and SF activity ($\mathrm{PC_2}$) increase significantly, both in luminosity and surface density.  This is a natural consequence of all parameters contributing to $\mathrm{PC_1}$ and $\mathrm{PC_2}$ increasing from Type B to Type C/D (see table \ref{avg_param}).  Although the increase of each parameters contributing to $\mathrm{PC_1}$ and $\mathrm{PC_2}$ are insignificant or small considering their standard deviations, the PCs with these parameters combined result in the distinct distribution of ISM content ($\mathrm{PC_1}$) and SF activity ($\mathrm{PC_2}$) between Type B and Type C/D.

  Nominally, an increase in $\log L'_\mathrm{3-2}$ can be attained either from collapse (increase in density) or heating of diffuse gas traced in $^{12}\mathrm{CO}(J=1-0)$.  $r_\mathrm{31}$ indeed increases significantly from Type B to C \citep{miura12}.  The increase in $r_\mathrm{31}$ could be due to gas heating by the increased star formation event, or alternatively, could indicate that ISM is being aggregated from the diffuse material.  ISM build-up may explain the 60\% increase in dust mass from Type B to C,  although a part of the increased dust mass may be due to enrichment by red supergiants and/or the death of massive stars occuring during the $\sim$10 Myr lifetime of Type B clouds \citep{dunne03, hernandez06, eldridge11}.

Clouds with larger mass tend to have higher $r_{31}$, higher $SFR$ and increased star formation efficiency \citep{onodera12, miura14}, resulting in higher $\mathrm{PC_1}$ and $\mathrm{PC_2}$.  Figure 1 (right) shows that the density of ISM content and SF activity increase as well.  This is because the average size of the clouds determined from $^{12}\mathrm{CO(J=3-2)}$, given in table \ref{avg_param} (taken from table 6 in \citet{miura12}) , indicate no size difference between Type B and C clouds.  A slight increase in cloud size for Type D may explain the slight surface density decrease in $\mathrm{PC_1}$ and $\mathrm{PC_2}$ from Type C to D.  

\begin{table*}
\caption{Average physical parameters of the sample GMCs}\label{avg_param}
\begin{center}
\begin{tabular}{llrrrr}
\hline
 &  & Type B & Type C & Type D & All \\  \hline
$\log L'_\mathrm{1-0}$ & $\mathrm{K\ km\ s^{-1}\ pc^2}$ & 4.10$\pm$0.41 &  4.14$\pm$0.26 & 4.22$\pm$0.27 & 4.16$\pm$0.29 \\
$\log L'_\mathrm{3-2}$ & $\mathrm{K\ km\ s^{-1}\ pc^2}$ & 4.56$\pm$0.42 &  4.77$\pm$0.29 & 4.83$\pm$0.26 & 4.75$\pm$0.31  \\
$\log M_\mathrm{d}$ & $\mathrm{M_\odot}$           & 2.52$\pm$0.33 &  2.74$\pm$0.24 & 2.78$\pm$0.21 & 2.72$\pm$0.26 \\
$\log M_*$ & $\mathrm{M_\odot}$                           & 4.89$\pm$0.31 &  5.51$\pm$0.32  & 5.40$\pm$0.46 & 5.37$\pm$0.43 \\
$\log SFR$ & $\mathrm{M_\odot yr^{-1}}$               &-4.03$\pm$0.64 & -3.03$\pm$0.61 & -3.18$\pm$0.46 & -3.25$\pm$0.67    \\
\hline
$\log \Sigma_{L^{\prime} 1-0}$ & $\mathrm{K\ km\ s^{-1}}$ &  0.59$\pm$0.35 &   0.63$\pm$0.30 &  0.57$\pm$0.24 & 0.61$\pm$0.29  \\
$\log \Sigma_{L^{\prime} 3-2}$ & $\mathrm{K\ km\ s^{-1}}$ &  1.05$\pm$0.30 &   1.26$\pm$0.33 &  1.18$\pm$0.28 & 1.20$\pm$0.32 \\
$\log \Sigma_{\mathrm{d}}$ & $\mathrm{M_\odot pc^{-2}}$              & -0.99$\pm$0.19 &  -0.76$\pm$0.22 & -0.87$\pm$0.18 & -0.83$\pm$0.22 \\
$\log \Sigma_{*}$ & $\mathrm{M_\odot pc^{-2}}$                              &  1.38$\pm$0.40 &   2.00$\pm$0.31  &  1.75$\pm$0.26 & 1.82$\pm$0.39  \\
$\log \Sigma_{SFR}$ & $\mathrm{M_\odot yr^{-1}\ pc^{-2}}$            & -7.54$\pm$0.51 &  -6.53$\pm$0.62 &  -6.83$\pm$0.60 & -6.80$\pm$0.69  \\
\hline
cloud radius & pc & 33 $\pm$ 10 & 33 $\pm$ 7 & 39 $\pm$ 10 \\ \hline
\end{tabular}
\end{center}
\end{table*}

\begin{figure*}[ht]
\begin{center}
\begin{tabular}{c}
  \begin{minipage}{0.33\hsize}
    \begin{center}
\includegraphics[clip,width=6.3cm]{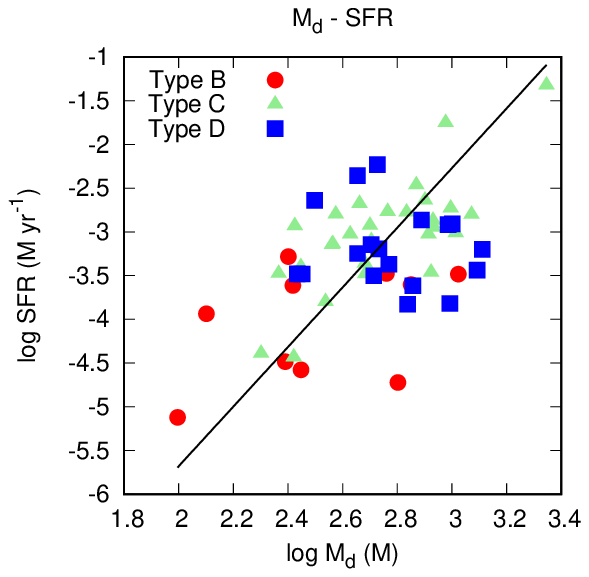}\label{co10-sfr-lum}
    \end{center}
  \end{minipage}
  \begin{minipage}{0.33\hsize}
    \begin{center}
\includegraphics[clip,width=6.3cm]{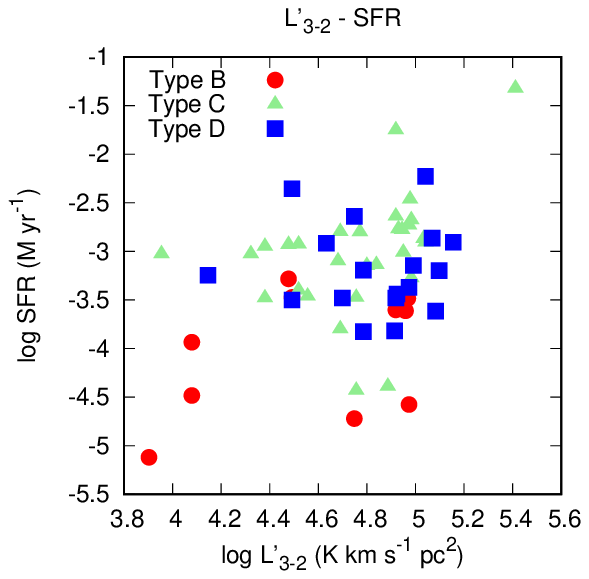}\label{co10-sfr-lum}
    \end{center}
  \end{minipage}
  \begin{minipage}{0.33\hsize}
    \begin{center}
\includegraphics[clip,width=6.3cm]{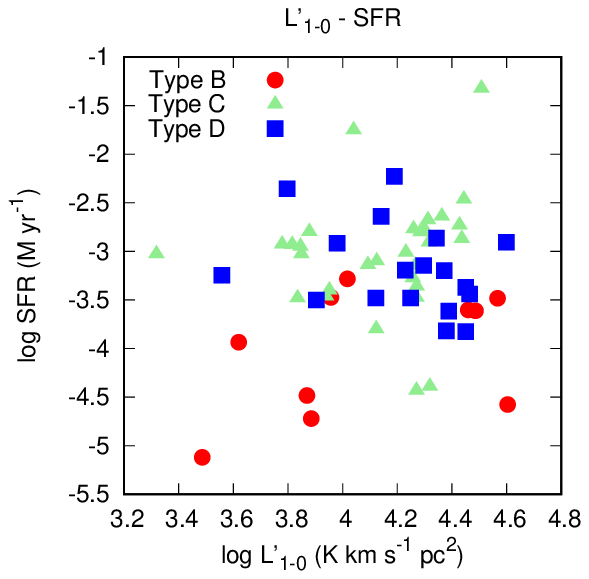}\label{co10-sfr-lum}
    \end{center}
  \end{minipage}
\end{tabular}

\begin{tabular}{c}
  \begin{minipage}{0.33\hsize}
    \begin{center}
\includegraphics[clip,width=6.3cm]{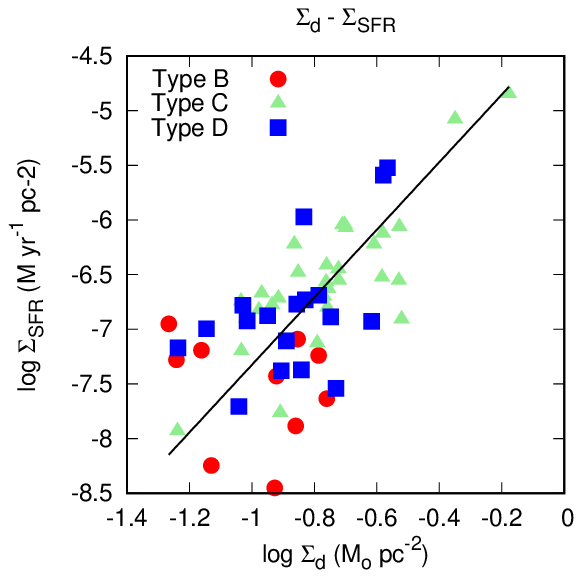}\label{co10-sfr-dens}
    \end{center}
  \end{minipage}
  \begin{minipage}{0.33\hsize}
    \begin{center}
\includegraphics[clip,width=6.3cm]{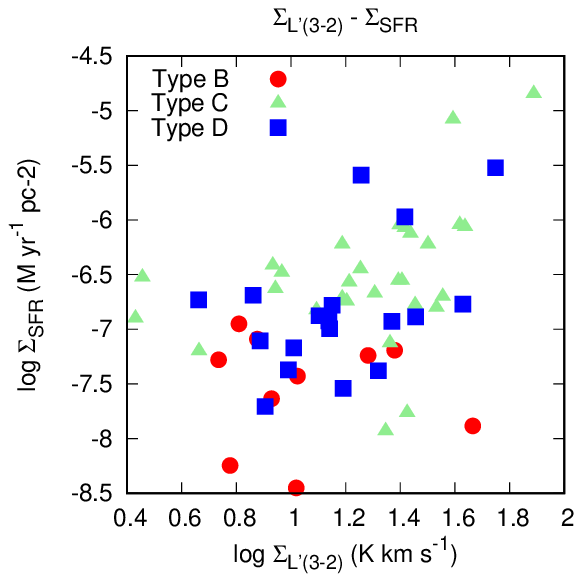}\label{co10-sfr-dens}
    \end{center}
  \end{minipage}
  \begin{minipage}{0.33\hsize}
    \begin{center}
\includegraphics[clip,width=6.3cm]{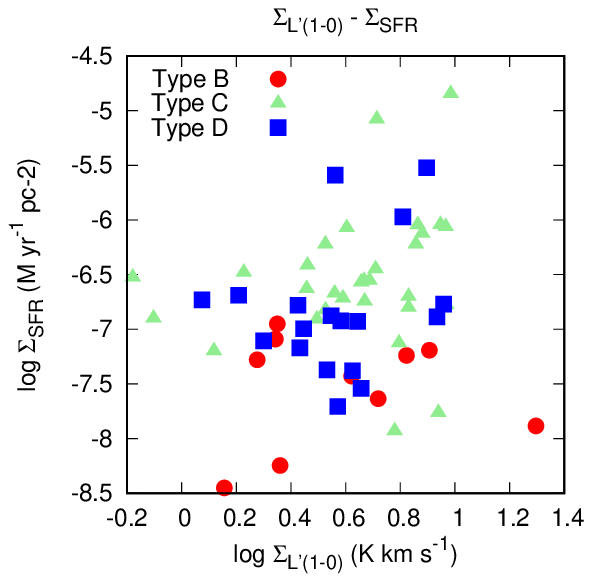}\label{co10-sfr-dens}
    \end{center}
  \end{minipage}
\end{tabular}
\caption{The $SFR$ as a function of dust (left), $^{12}\mathrm{CO}(J=3-2)$ (middle), and $^{12}\mathrm{CO}(J=1-0)$ (right).  Top panels are for values in luminosity, bottom for surface densities.  Symbols are coded as in figure 1.  The black lines in the left panels indicate equations 7 and 9, obtained by setting $PC^{\prime}_4=0$.}
\end{center}
\end{figure*}\label{sk_dens}

\subsection{Dynamical Equilibrium of the Interstellar and Stellar Phases}

Although parameters which contribute weakly (i.e., with coefficients smaller than 0.45 in tables \ref{coeff_lum} and \ref{coeff_surface}) are dropped from the analyses above, it is worth noting some relations with these weakly contributing parameters to compare with previous studies.  From tables \ref{coeff_lum} and \ref{coeff_surface}, it can be seen that $\mathrm{PC_5}$ is almost completely dominated by contributions from $L^{\prime}_\mathrm{3-2}$ and $L^{\prime}_\mathrm{1-0}$.  However, the effect of $\log SFR$ is visible when the three parameters are seen in three dimensional space, where the two CO luminosities and the $SFR$ comprise a plane expressed by 
\begin{eqnarray}
\log L^{\prime}_\mathrm{3-2}  & = & 1.0\log L^{\prime}_\mathrm{1-0} + 0.12\log SFR + 2.3  \quad (\mathrm{PC^{\prime}_5=0})  \\
\log L^{\prime}_\mathrm{3-2}  & = & (0.93 \pm 0.05)\log L^{\prime}_\mathrm{1-0} \nonumber \\
& + & (0.12\pm 0.02)\log SFR \nonumber \\
& + & 1.3 \pm 0.2 \quad (\mathrm{least\ squares\ fitting} )
\end{eqnarray}
for luminosities, and
\begin{eqnarray}
\log \Sigma_{L^{\prime} 3-2} & = & 1.0\log \Sigma_{L^{\prime} 1-0} -  0.10\log \Sigma_{SFR}  + 2.9  \quad (\mathrm{PC^{\prime}_5=0}) \\
\log \Sigma_{L^{\prime} 3-2} & = & (0.94\pm 0.05)\log \Sigma_{L^{\prime} 1-0}  \nonumber \\
& - & (0.12\pm 0.02)\log \Sigma_{SFR}  \nonumber \\
& + & (1.4\pm 0.2)  \quad  (\mathrm{least\ squares\ fitting})
\end{eqnarray}
in case of surface densities, both with scatter of 0.2 dex (figure 3 left panel).  The equations are derived in the same way as in equations 7 and 9 using table \ref{avg_param}.
The near-constant ratio of the CO intensities slightly modulated by the  $SFR$, is consistent with results by \citet{onodera12} which find that more massive molecular clouds with higher $SFR$ tend to have higher $r_{31}$.  The small but present effect of the SFR to the CO ratio is visible because the contribution from other parameters (stars and dust) are effectively zero.  The local (GMC scale) SK plots in figure 2 (bottom middle and bottom right) are two-dimensional projections of the plane defined in equation 13.

\begin{figure*}[ht]\label{fig_pca}
\begin{center}
\begin{tabular}{c}
  \begin{minipage}{0.5\hsize}
    \begin{center}
      \includegraphics[clip,width=8.7cm]{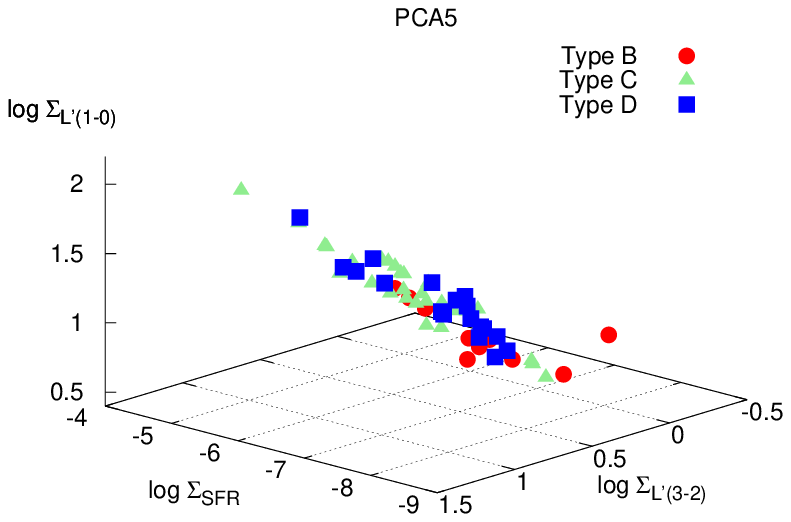}\label{pca5-3d}

    \end{center}
  \end{minipage}
  \begin{minipage}{0.5\hsize}
    \begin{center}
      \includegraphics[clip,width=8.7cm]{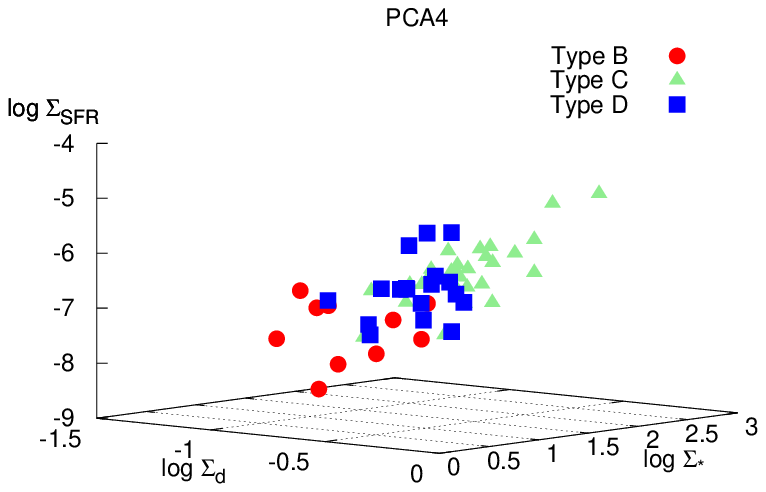}\label{pca4-3d}

    \end{center}
  \end{minipage}
 \end{tabular}
\caption{The plane defined by equation 13 (left panel) and equation 15 (right panel).  Figures are oriented so that the best fit plane is seen observed edge on.  Symbols are coded as in figure 1.  }
\end{center}
\end{figure*}

A number of previous studies have pointed out the possible contribution of stellar surface density to star formation \citep{blitz06, shi11, rahmani16, dib17}.  In this study also, the stellar mass contributes mildly to $\mathrm{PC_4}$ in case of surface densities.  The plane formed by $\log \Sigma_{SFR}$, $\log \Sigma_{\mathrm{d}}$ and $\log \Sigma_{*}$ are
\begin{eqnarray}
\log \Sigma_{SFR} & = & 3.1\log \Sigma_{\mathrm{d}} + 0.67\log \Sigma_{*} - 5.5 \quad (\mathrm{PC^{\prime}_4=0}) \\
\log \Sigma_{SFR} & = & (2.1 \pm 0.2)\log \Sigma_{\mathrm{d}} \nonumber  \\
& + & (0.5 \pm 0.2)\log \Sigma_{*}  \nonumber \\
& - & (5.9 \pm 0.4) \quad \mathrm{least\ squares\ fitting}
\end{eqnarray}
with a scatter of 0.4 dex (figure 3 right panel).  As discussed in section 5.1, if dust indeed traces molecular gas better than CO lines at scales of individual molecular clouds, the relation given in equation 15 can be interpreted as an extended form of the SK relation that connects $SFR$ with gaseous and stellar content, valid at GMC scales.  The power law index of the stellar term (0.67 from PCA, 0.5 from least squares fitting) is roughly consistent with that in recent literature (0.36 in \cite{shi11}, 0.74 in \cite{rahmani16}) and theoretical expectations in the case of star formation triggered by two-fluid instabilities \citep{dib17}.  \citet{blitz06} show that for pressure regulated star formation, the stellar contribution has an index of roughly 0.5 and molecular gas density has an index of 2.0 (where equations 15 and 16 indicate that the dust term has an index of 2 to 3) .  It should be noted, however, that stellar contribution can be seen only for surface densities ($(-0.26)^2 \sim 6\%$) but is $0\%$ in case of luminosities.  This difference may either indicate that the extended star formation law is indeed sensitive to density, or simply that the appearance of a stellar contribution is due to small number statistics in this study.

 An interesting feature of equations 13 and 15 is that the relation holds for clouds regardless of evolutionary stage.  GMC types B, C and D mark evolution timescales of 10 Myr, over which the conversion from gas to stars proceed in tandem with feedback processes from stars, resulting in significant changes in distribution of ISM and stars.  Thus, one may expect that clouds of different types would lie on a different relation (provided that any relations exist at this scale).  The fact that GMCs of different types spanning $\sim$40 Myr lie on the same planes shown in figure 3 indicate that equations 13 and 15 act as boundary conditions for how physical parameters can change, i.e., can be interpreted as a kind of equation of state describing the dynamical equilibrium between the ISM and stellar phases.  

\subsection{Future Work}
This paper presents a case study on a late type galaxy which is relatively metal poor.  There are a number of other physical parameters which are expected to relate to star formation.  In particular, metallicity and dynamical parameters (velocity dispersion and larger scale shear) are not included in the PCA analysis presented here.  Further studies should include a significant increase in sample size, spanning a broader range of quantified and catalogued characteristics.






\begin{ack}
The authors thank the referee for helpful comments which improved the paper.  S. Komugi is financially supported by the Japan Society for the Promotion of Science (JSPS) KAKENHI (25800114, 15H02074).
\end{ack}


\end{document}